\documentclass[reprint,amsmath,amssymb,prl,superscriptaddress,%
longbibliography]{revtex4-1}

\usepackage{graphicx}
\usepackage{bm}
\usepackage[utf8]{inputenc}
\usepackage{color}
\definecolor{goodgreen}{rgb}{0.1,0.5,0}
\definecolor{goodred}{rgb}{0.7,0,0}
\definecolor{goodblue}{rgb}{0,0,0.8}
\usepackage[colorlinks,urlcolor=blue,citecolor=goodgreen,linkcolor=goodred]%
{hyperref}

\usepackage{mathptmx}
\usepackage[scaled=.90]{helvet}
\usepackage{courier}

\newcommand{\vg}{\ensuremath{V_{\text{g}}}}
\newcommand{\vsd}{\ensuremath{V_{\text{sd}}}}
\newcommand{\Nel}{\ensuremath{N_{\text{el}}}}
\newcommand{\didv}{\ensuremath{\text{d}I/\text{d}\vsd}}
\newcommand{\un}[1]{\ensuremath{\,\text{#1}}}

\begin{document}

\title{Carbon Nanotube Millikelvin Transport and Nanomechanics}

\author{K. J. G. Götz}
\affiliation{Institute for Experimental and Applied Physics,
University of Regensburg, Universit\"{a}tsstra{\ss}e 31, 93053 Regensburg,
Germany}
\author{F. J. Schupp}
\affiliation{Institute for Experimental and Applied Physics,
University of Regensburg, Universit\"{a}tsstra{\ss}e 31, 93053 Regensburg,
Germany}
\affiliation{Department of Physics, Princeton University, Princeton, New Jersey
08544, USA}
\author{A. K. Hüttel}
\affiliation{Institute for Experimental and Applied Physics,
University of Regensburg, Universit\"{a}tsstra{\ss}e 31, 93053 Regensburg,
Germany}
\email{andreas.huettel@ur.de}

\begin{abstract}
Single wall carbon nanotubes cooled to cryogenic temperatures are outstanding
electronic as well as nano-electromechanical model systems. To probe a largely
unperturbed system, we measure a suspended carbon-nanotube device where the
nanotube is grown in the last fabrication step, thus avoiding damage and
residues from subsequent processing. In this ultra-clean device, we observe
the transport spectrum and its interaction with nano-electromechanics over a
wide gate voltage range and thereby over a wide range of coupling parameters
between the quantum dot and the contact electrodes.
\end{abstract}

\maketitle

\section{Introduction}

Both the electronic and the nano-electromechanical low-temperature properties 
of single-wall carbon nanotubes have attracted significant research over the
past decades, resulting in a large spectrum of publications
\cite{rmp-laird-2015}. In particular ``ultraclean'', as grown carbon nanotubes
are prototypical single-electron devices \cite{nmat-cao-2005}, where the
unperturbed transport spectrum reveals a wide range of phenomena originating 
from the band structure
\cite{nature-kuemmeth-2008,nnano-steele-2009,nnano-pei-2012,brokensu4} 
and the interaction between the charge carriers
\cite{nphys-deshpande-2008,nature-hamo-2016,arxiv-shapir-2018,prl-island-2018}.
At the same time, single-wall carbon nanotubes at millikelvin temperatures are
excellent mechanical resonators
\cite{nature-sazonova-2004,nl-witkamp-2006,highq}, with resonance frequencies
up to $39\un{GHz}$ \cite{nl-laird-2012} and quality factors up to $5\times
10^6$ \cite{nnano-moser-2014}. In suspended carbon nanotube devices, the 
electrostatic forces acting on single electrons dominate the mechanical 
behaviour, and electronic transport is subject to strong coupling between 
single electron tunneling and vibrational motion \cite{strongcoupling,highqset,%
science-lassagne-2009}.

Typically, reports focus on selected aspects of the behaviour of carbon
nanotubes, possibly across several devices. While this is a suitable approach
for in-depth analysis of the observed phenomena, it is challenging to obtain a 
complete picture of the behaviour of one single wall nanotube device across all 
regimes of coupling to the leads. Here, we present low-temperature measurement 
data collected on a single device. We follow both electronic and
nano-electromechanical phenomena from the highly transparent transport through
hole states, across the band gap to strong Coulomb blockade, and finally towards
larger electron numbers, where the tunnel barriers become increasingly 
transparent with higher order transport processes dominating the current.

The schematic device geometry is sketched in
\begin{figure}[t]
\centering
\includegraphics[width=\columnwidth]{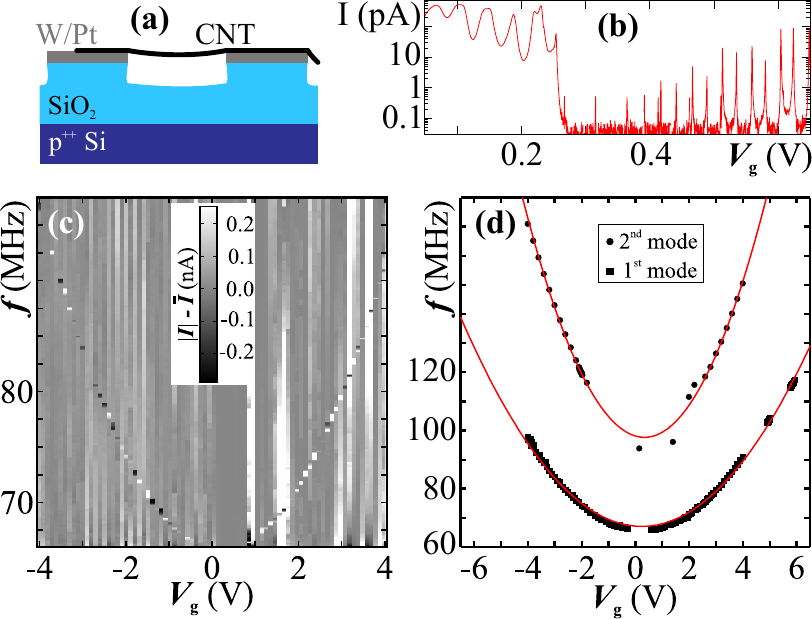}
\caption{
(a) Sketch of the geometry of the measured device. 
(b) Current $I(\vg)$ at $\vsd = 20\un{$\mu$V}$ around the electronic band gap. 
Hole transport for $\vg<0.3\un{V}$, band gap, and electron transport for 
$\vg>0.3\un{V}$ with an electronic shell pattern can be clearly discerned.
(c) Measured dc current as function of gate voltage \vg\ and driving rf 
frequency $f$,  $I(f, \vg) - \overline{I}(\vg)$, with the average current per 
frequency trace subtracted. The transversal vibration resonance becomes visible 
as a sequence of peaks or dips \cite{highq}. $\vsd=-0.1\un{mV}$, nominal rf
generator power $P=2\un{dBm}$.
(d) Extracted resonance frequency $f_0(\vg)$ for the fundamental vibration mode 
of (c) and a higher mode; the solid lines are parabolic fits.
\label{fig:device}} 
\end{figure}
Fig.~\ref{fig:device}(a). The device fabrication follows the approach of 
\cite{nmat-cao-2005}. We start with a highly p++ doped silicon wafer having a 
500nm dry-grown thermal surface oxide. The contact electrodes, with a
bilayer of 10nm tungsten and 40nm platinum, are fabricated using electron beam 
lithography, metal evaporation, and lift-off. Subsequently, the silicon oxide 
surface is 
anisotropically etched by 100nm to deepen the trenches between the electrodes. 
Nanotube growth catalyst is locally deposited via an additional lithography 
step, drop-casting, and lift-off \cite{nature-kong-1998}. In the final 
fabrication step, the carbon nanotubes are grown in situ by chemical vapour 
deposition \cite{nature-kong-1998}. The distance between the contact electrodes 
is $L = 1.2\,\mu\text{m}$, providing a lower boundary for the length of the 
active nanotube segment.

A measurement of the low-bias ($\vsd = 20\un{$\mu$V}$) current through the 
device at base temperature $T \le 25\un{mK}$ of the dilution refrigerator is
plotted in Fig.~\ref{fig:device}(b). On the hole conduction side, i.e. for $\vg 
< 0.3\un{V}$, the device is highly transparent, while the electron conduction 
side, $\vg > 0.3\un{V}$, initially displays strong Coulomb blockade. This 
indicates that the actual contacts between nanotube and platinum electrodes are
transparent. The opaque tunnel barriers of the quantum dot at low electron 
number are given by the extended p-n junctions between the electrostatically 
n-doped central nanotube segment and the nanotube segments close to the leads 
\cite{apl-park-2001,nature-kuemmeth-2008,highfield}.

Figure~\ref{fig:device}(c) shows a measurement of the transversal vibrational
resonance of the carbon nanotube \cite{nature-sazonova-2004,%
nl-witkamp-2006}. Here, the detection scheme of \cite{highq,strongcoupling,%
magdamping} is used: an antenna several millimeters from the device radiatively
introduces a MHz signal, and the resulting driven vibration at mechanical
resonance leads to a peak or dip in detected dc-current due to the modulated
displacement of the nanotube and the resulting modulation in gate capacitance.
We observe a strong gate voltage dependence of the resonance frequency
$f_0(\vg)$ due to the changing tension in the nanotube induced by the pull of
the gate voltage. The resonance from Fig.~\ref{fig:device}(c) and an additional
second mode at higher frequency are extracted in Fig.~\ref{fig:device}(d). In
both cases, a parabolic fit provides a good (though strongly simplified) 
approximation of the gate voltage dependence in the measured range (cf.
\cite{nl-witkamp-2006,pssb-poot-2007} for a more detailed model).

\begin{figure}[t]
\centering
\includegraphics[width=\columnwidth]{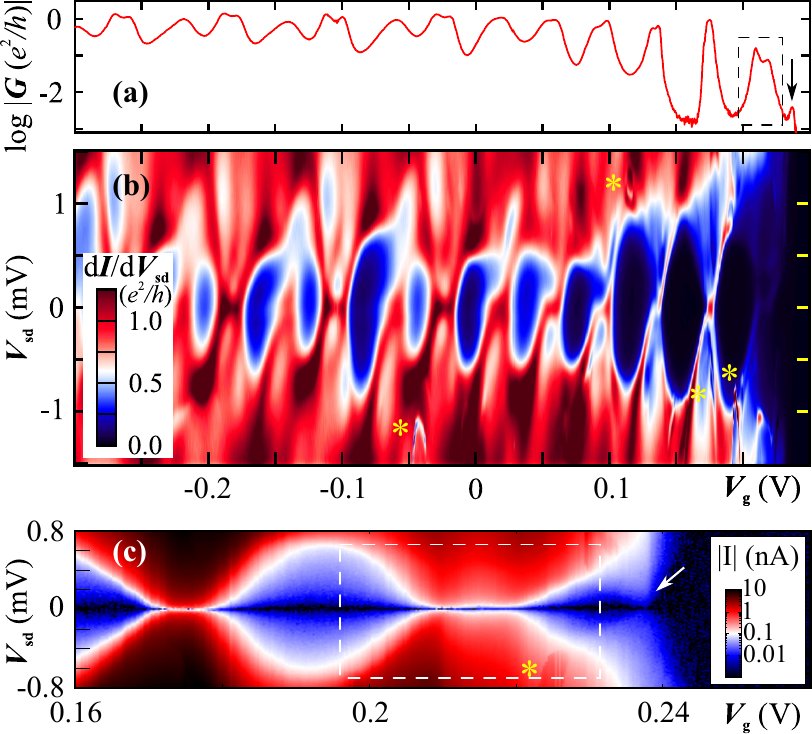}
\caption{
Transport spectrum of the few-hole parameter region of the device: (a)
zero-bias differential conductance $\didv(\vg)$ and (b) stability diagram
$\didv(\vg, \vsd)$ at identical \vg-axis. Zero bias ridges of enhanced
conductance indicate the presence of the Kondo effect. (c) Detail zoom of the
band gap edge, now plotting the absolute value of the measured current
$\left|I(\vg)\right|$. Yellow asterisks mark mechanical instability features, 
see the text.
\label{fig:holes}} 
\end{figure}
\section{Hole transport region}

At gate voltages $\vg<0.3\un{V}$, current is carried by valence band states, 
and the quantum dot is charged with an increasing number 
of holes for decreasing gate voltage. Due to the absence of abovementioned p-n 
barriers, the device displays strong coupling to the metal contacts. This is 
illustrated in Fig.~\ref{fig:holes}; note that while the line plot of 
Fig.~\ref{fig:holes}(a) is logarithmic, the color scale of 
Fig.~\ref{fig:holes}(b) is linear. The pronounced sequence of conductance
oscillations at zero bias turns, upon closer observation, out to be a sequence
of Kondo conductance ridges \cite{nature-goldhabergordon-1998,%
nature-nygaard-2000,nature-jarillo-2005,brokensu4} corresponding each to the
addition of two holes. This is particluarly obvious around
$\vg=-0.03\un{V}$, $-0.11\un{V}$, $-0.19\un{V}$, where a clear Kondo zero bias
anomaly of conductance emerges. The yellow asterisks in Fig.~\ref{fig:holes} 
mark characteristic lobe-shaped features with sharp edges in the data which 
correspond to mechanical instability and vibrational feedback phenomena in 
transport \cite{prb-usmani-2007,strongcoupling,magdamping,heliumdamping}.

Figure~\ref{fig:holes}(c) zooms in on the band gap edge; here we plot the
absolute value of the dc current $\left| I(\vg,\vsd)\right|$ in logarithmic
scale since this allows us to resolve smaller signals. The feature at $\vg =
0.22\un{V}$ turns out to be a clear double-peak, i.e., two nearly merged
Coulomb oscillations, consistent with the curve shape of
Fig.~\ref{fig:holes}(a) [see the dashed boxes in both figures]. Since the Kondo
ridges are associated with odd electron or hole numbers, this means that at
$\vg\approx 0.23\un{V}$ the nanotube is charged with an even number of holes.
This leaves us with a conundrum about the nature of the structure at $\vg
\approx 0.24\un{V}$, also visible as a single sharp peak in
Fig.~\ref{fig:holes}(a) and marked in both figures with an arrow. Two
explanations are possible: either this is a single Coulomb oscillation and an
additional oscillation cannot be resolved due to too small tunnel rates, or
this peak is already a merger of two oscillations. The former explanation seems
unlikely since, even at high bias, no further structures in current or
conductance are visible at the hole side of the band gap region. Measurements
at larger temperature and at finite magnetic field may provide additional
information.

\begin{figure}[t]
\centering
\includegraphics[width=\columnwidth]{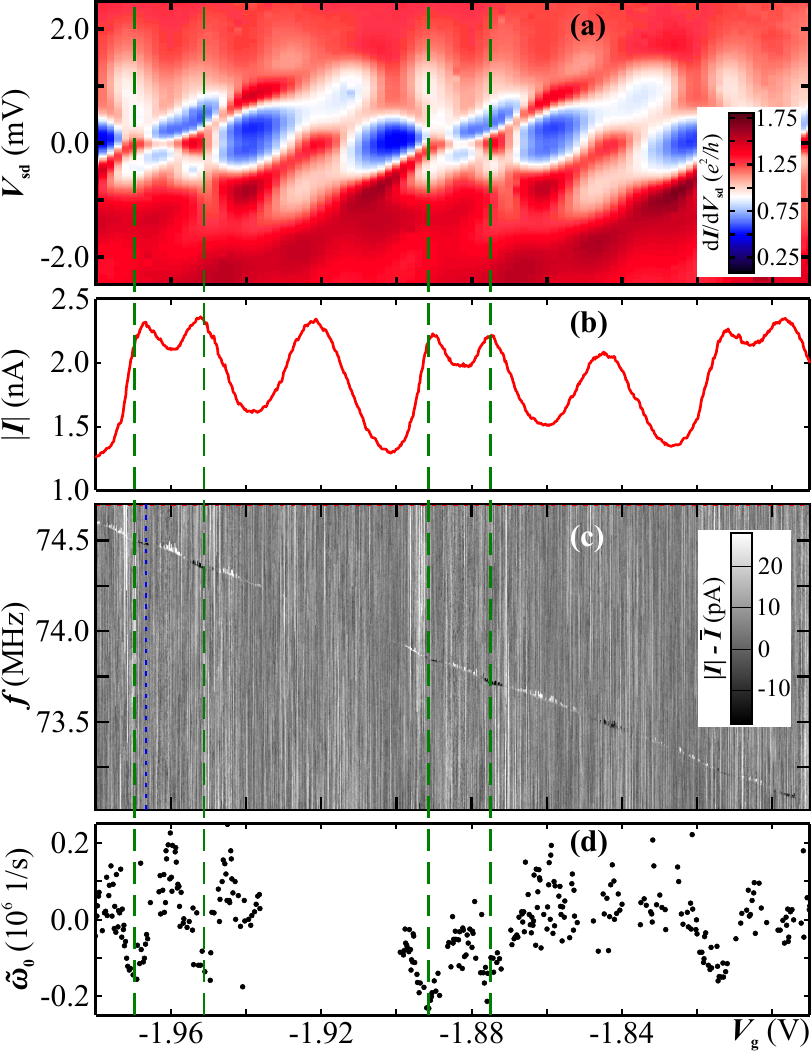}
\caption{Behaviour at large hole number, in the transition region between Kondo
transport and Fabry-P\'erot interference. 
(a) Stability diagram showing $\didv(\vg,\vsd)$. Again zero-bias ridges of 
enhanced conductance clearly indicate the Kondo
effect. 
(b)--(d) Mechanical resonance frequency detection in the same gate
voltage region: (b) off-resonant current $I(\vg)$ for $\vsd=-0.1\un{mV}$, (c)
current as function of gate voltage \vg\ and driving rf frequency $f$, $I(f,
\vg)-\overline{I}(\vg)$, with the average current per frequency trace
subtracted; $\vsd=-0.1\un{mV}$, nominal rf driving signal power
$P=-13.7\un{dBm}$. (d) Resonance (angular) frequency $\tilde{\omega}_0(\vg) = 
2\pi f_0(\vg) - (a\vg+b)$ extracted from (c), with a term linear in \vg\ 
subtracted; see also \cite{strongcoupling,kondocharge,prb-meerwaldt-2012}.
\label{fig:holemech}} 
\end{figure}
A stability diagram at larger hole number, where the interfaces between the 
carbon nanotube and its contact electrodes become increasingly transparent 
towards Fabry-P\'erot conduction \cite{nature-liang-2001,fabryperot}, is plotted 
in Fig.~\ref{fig:holemech}(a). In absence of the (SU(2)) Kondo effect at even 
electron number, elastic co-tunneling enables large conductance in nominal 
Coulomb blockade regions \cite{prl-defranceschi-2001}. For odd electron 
numbers, Coulomb blockade is suppressed at low bias, and, as already observed 
in Fig.~\ref{fig:holes}, several of the conductance peaks merge completely into 
Kondo ridges. This is also clearly visible in the line plot of 
Fig.~\ref{fig:holemech}(b), displaying the current at $\vsd = -0.1\un{mV}$.

The subseqent panel, Fig.~\ref{fig:holemech}(c), shows the result of 
mechanical resonance detection in the same gate voltage region, using the same 
measurement and plotting scheme as Fig.~\ref{fig:device}(c), though at much 
lower driving power (nominally $P=-13.7\un{dBm}$). The fundamental vibration
mode becomes clearly visible as a diagonal feature. In the plotted gate voltage
region it displays a near-linear dependence on the gate voltage \vg,
qualitatively different from the observations of, e.g., \cite{strongcoupling,%
prb-meerwaldt-2012,nphys-benyamini-2014,nl-hakkinen-2015}, where each Coulomb 
oscillation causes a strong dip in frequency due to changing electrostatic 
forces in response to single charge tunneling at the charge degeneracy 
points. This indicates that, due to increased tunnel rates and lifetime 
broadening of the involved quantum levels, Coulomb blockade is gradually 
lost \cite{kondocharge} and that the nanotube more and more resembles a 
metallic beam. However, a more detailed analysis (Fig.~\ref{fig:holemech}(d)), 
where the resonance frequency $f_0(\vg)$ has been extracted and a linear 
contribution has been removed, still allows to identify some of the 
characteristic oscillations of $\tilde{\omega}_0(\vg)$, corresponding to the 
addition of single holes to the system \cite{strongcoupling,kondocharge}; 
selected features are marked in the figure with vertical dashed lines.

For even larger negative gate voltages, our device unambiguously enters the
Fabry-P\'erot regime, where weak scattering of the electronic wavefunctions at
the contacts causes quantum interference \cite{nature-liang-2001}. A detailed 
discussion of the properties of the device in this parameter region can be 
found in \cite{fabryperot}.

\begin{figure}[t]
\centering
\includegraphics[width=\columnwidth]{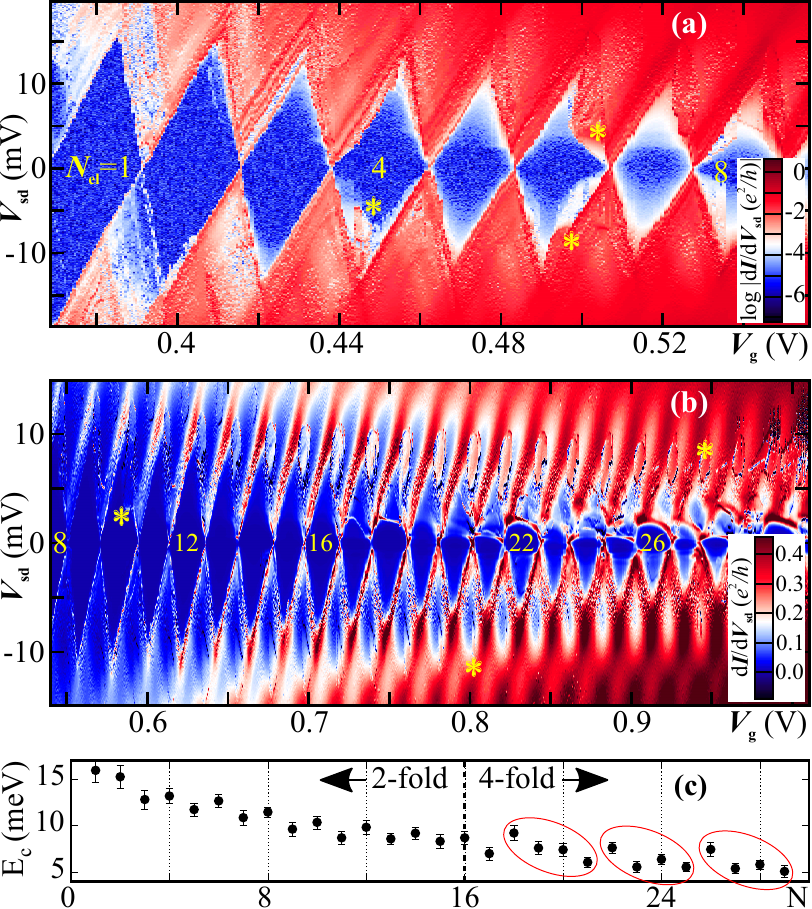}
\caption{
Stability diagrams showing $\didv(\vg,\vsd)$ of the strong Coulomb blockade 
parameter region at low electron number: 
(a) $1 \le \Nel \le 8$, logarithmic color scale, and 
(b) $8 \le \Nel \le 30$, linear color scale. Yellow asterisks mark (some of 
the) mechanical instability features. 
(c) Charging energy as function of electron number \Nel.
\label{fig:strongcb}} 
\end{figure}
\section{Strong Coulomb blockade}

The differential conductance measurements of Fig.~\ref{fig:strongcb} cover the
parameter region adjacent to the bandgap towards positive gate voltage, where 
few electrons are trapped in the carbon nanotube. Fig.~\ref{fig:strongcb}(a) 
shows the $1 \le \Nel \le 8$ region in logarithmic color scale. The 
diamond-shaped regions of Coulomb blockade and fixed number \Nel\ of trapped 
electrons are clearly visible. Additionally, we observe multiple conductance 
lines in the single electron tunneling regions corresponding to excited quantum 
states. Already in the Coulomb blockade region with $\Nel=4$, further 
lobe-shaped features appear, becoming much stronger at $\Nel=6$ (see the yellow 
asterisks in the figure). These again correspond to vibrational feedback 
phenomena typical for clean and suspended carbon nanotube devices at 
millikelvin temperatures
\cite{prb-usmani-2007,strongcoupling,magdamping,heliumdamping}.

Fig.~\ref{fig:strongcb}(b) continues the plot of Fig.~\ref{fig:strongcb}(a) 
towards larger electron numbers $8 \le \Nel \le 30$, now in linear color scale. 
The lobe-shaped instability regions now occur repetitively for every Coulomb 
oscillation. In addition, near-horizontal features in Coulomb blockade, which 
can be associated with inelastic cotunneling \cite{prl-defranceschi-2001} and 
the non-equilibrium Kondo effect 
\cite{nature-nygaard-2000,splitkondo,brokensu4}, become increasingly prominent.

A plot of the charging energy $E_c(\Nel)$ as function of electron number \Nel\ 
is provided in Fig.~\ref{fig:strongcb}(c). It displays the overall
decrease of $E_c$ with \Nel, typical for this type of device, in addition to 
"shell-effects" from subsequent filling of quantum levels analogous to shells in 
atomic physics. Interestingly, the shell-filling results in a two-fold pattern 
for $\Nel<16$, while for $\Nel>16$ a fourfold peak grouping with the largest 
charging energy at $4 n + 2$ is observed. Electron-electron interaction effects 
are a likely cause of this phenomenon \cite{nphys-deshpande-2008}, requiring 
further analysis and potentially additional measurements in magnetic fields.

\begin{figure}[b]
\centering
\includegraphics[width=\columnwidth]{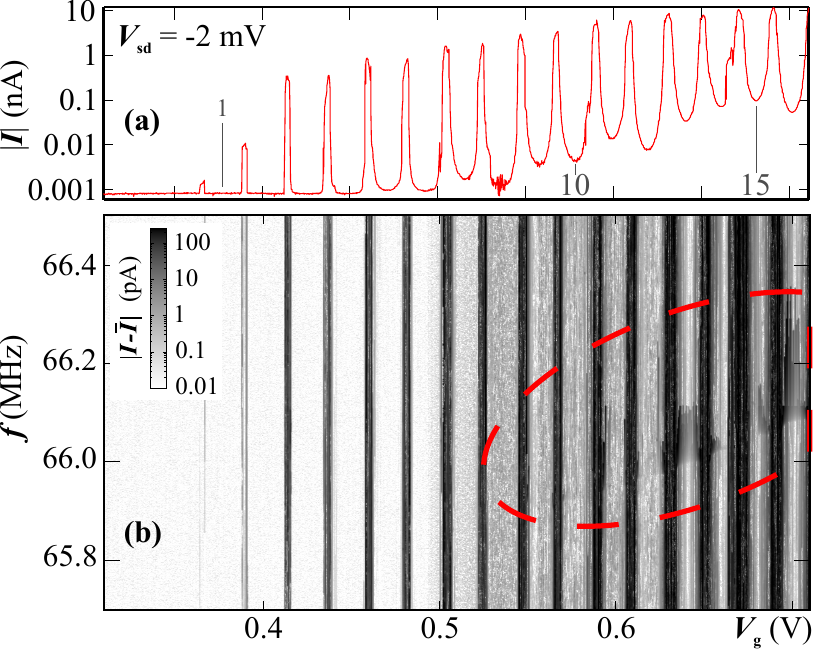}
\caption{
(a) dc current $I(\vg)$ at $\vsd=-2\un{mV}$ for off-resonant rf driving signal; 
trace cut from the raw data of (b). (b) Mechanical resonance detection 
measurement at nominal rf power $P=-13.7\un{dBm}$; $\left| I(f, \vg) - 
\overline{I}(\vg)\right|$ in logarithmic scale. The resonance is indicated by a 
dashed ellipsoid.
\label{fig:strongcbmech}}
\end{figure}
Figure~\ref{fig:strongcbmech} illustrates the attempt to resolve the
transversal vibration resonance for decreasing electron numbers \Nel\ all 
the way to the electronic band gap. Even though a comparatively large bias of 
$\vsd=2\un{mV}$ (see Fig.~\ref{fig:strongcbmech}(a)) is applied, the resonance 
peaks only remain detectable for $\Nel \ge 9$, as marked by an ellipsiod in 
Fig.~\ref{fig:strongcbmech}(b). The peaks are far in the nonlinear response 
regime, displaying an abrupt edge similar to a Duffing oscillator response at 
increasing driving frequency. For $\Nel < 9$ the resonant response becomes too 
small to be detected. To be able to trace the transversal vibration resonance 
frequency across the low electron number region, and specifically also the 
electronic band gap, a different detection scheme is required.

\begin{figure}[t]
\centering
\includegraphics[width=\columnwidth]{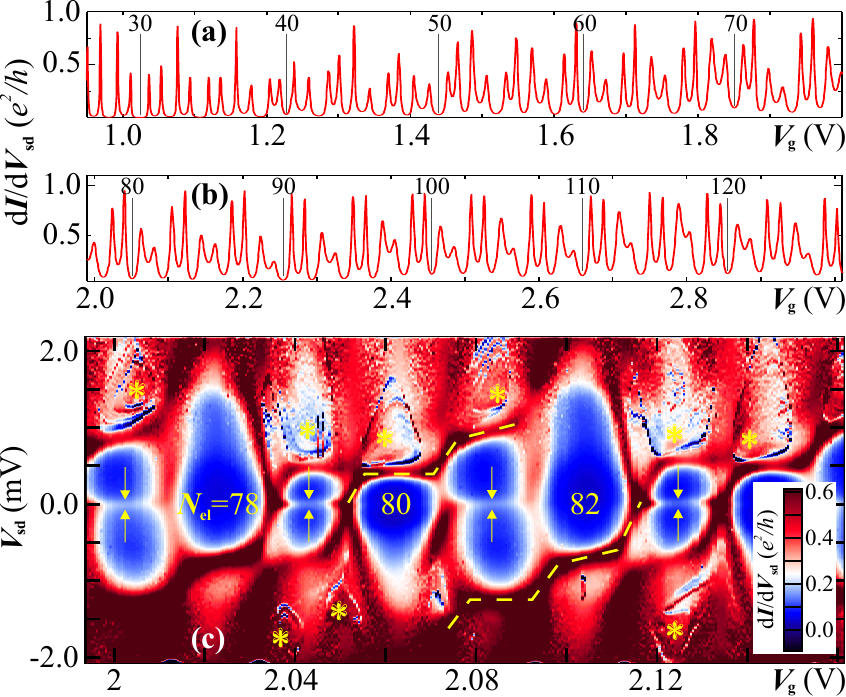}
\caption{
(a), (b) Zero-bias differential conductance $\didv(\vg)$ for increasingly
positive gate voltage. The number of trapped electrons, $27\le \Nel \le 77$ in
(a) and $77 \le \Nel \le 128$ in (b), is indicated in the line plots.
(c) Example stability diagram: differential conductance $\didv(\vg, \vsd)$ in
the region $77 \le \Nel \le 85$; linear color scale, cut off at $0.6 e^2/h$ for
better contrast.
\label{fig:manye}}
\end{figure}
\section{Large electron numbers}

With increasing positive gate voltage \vg\ and thereby increasing number of
trapped electrons \Nel, higher order tunneling processes become
dominant in the transport spectrum. This is immediately visible in the
differential conductance plots of Figs.~\ref{fig:manye}(a) and (b), where the
Kondo effect and (to a lesser degree) elastic cotunneling increase the signal.
While the conductance trace $\didv(\vg)$ retains an overall fourfold
repetitive pattern, corresponding to the fourfold occupation of each 
longitudinal momentum state in the nanotube confinement potential, the precise 
shape of the conductance oscillations continuously evolves towards Kondo ridges 
and suppressed Coulomb blockade.

An example stability diagram, plotting the differential conductance $\didv 
(\vg, \vsd)$ as function of gate voltage \vg\ and bias voltage \vsd\ in the 
region $77 \le \Nel \le 85$, is shown in Fig.~\ref{fig:manye}(c). A zero bias
conductance anomaly becomes clearly visible at odd electron number, though it 
is here still comparatively weak. An evaluation of the Kondo ridge width, 
following \cite{prb-kretinin-2012}, leads to typical Kondo temperatures of 
$1\un{K} \le T_\text{K} \le 2\un{K}$ at the center of the depicted odd-\Nel\ 
Coulomb blockade regions.

Much more dominant in Fig.~\ref{fig:manye}(c) are, however, different effects.
As marked by dashed lines in the figure, strong, discrete conductance 
resonances at finite bias pass stepwise through the Coulomb blockade regions. 
Going beyond inelastic cotunneling, which would lead to a conductance threshold
alone, these lines again correspond to nonequilibrium Kondo phenomena 
\cite{nature-nygaard-2000,splitkondo,brokensu4}. Similar ``stepwise'' resonance 
structures have been observed before and identified with an imbalance of the 
two tunnel barriers connecting a quantum dot to its leads 
\cite{prl-makarovski-2007}. Finally, regions of mechanical instability, marked 
with asterisks in Fig.~\ref{fig:manye}(c), now consistently occur at nearly 
every electron number for bias values above $\left| \vsd \right| \simeq 
0.5\un{mV}$ and strongly distort the transport spectrum.

\section{Conclusions}

While a clean and suspended carbon nanotube may seem like a comparatively 
simple object, its spectroscopy reveals an immense richness of phenomena in 
electronics, nanomechanics, and the coupling of both. The tunability of the 
tunnel couplings via the gate voltage allows access to a wide parameter range, 
from the electronic band gap and strong Coulomb blockade all the way to higher 
order tunneling phenomena and eventually Fabry-P\'erot interference. In nearly 
all of these regions nano-electromechanical phenomena can be identified. A 
driving signal makes it possible to detect the transversal vibration resonance 
frequency from low electron numbers all the way to the strong Kondo regime; 
even where charge quantization is nearly lost, a detailed analysis of the 
resonance frequency evolution still allows to characterize the electronic 
system via its quantum capacitance. Additionally, nano-electromechanical 
feedback effects dominate the current at finite bias, leading to strongly 
distorted transport spectra. The tunability of a single, clean device allows 
the direct comparison of parameter regimes \cite{ncomm-niklas-2016}, at known 
electron number and unchanged molecular structure.

\section*{Acknowledgments}

We would like to thank Ch. Strunk and D. Weiss for the use of experimental
facilities. The authors acknowledge funding by the Deutsche
Forschungsgemeinschaft via Emmy Noether grant Hu 1808/1, SFB 689, and SFB 1277.
The measurement data has been recorded using the
\href{https://www.labmeasurement.de/}{Lab::Measurement}
software package \cite{labmeasurement}.

\bibliography{paper}

\end{document}